\documentstyle[twoside,fleqn,npb,epsfig]{article}

\newcommand{\AmS}{{\protect\the\textfont2
  A\kern-.1667em\lower.5ex\hbox{M}\kern-.125emS}}

\title{Perturbative contribution to the {\it sin}$\phi$  
       asymmetry in inclusive $\pi^{+}$ electroproduction\thanks{Talk 
presented at the QCD'00 Euroconference, Montpellier, 6-13th July 2000. }
} 

\author{K.~A.~Oganessyan\address{INFN-Laboratori Nazionali di Frascati, 
via Enrico Fermi 40, I-00044 Frascati, Italy}\address{DESY, Notkestrasse 
85, 22603 Hamburg, Germany}\thanks{E-mail: kogan@hermes.desy.de}
\thanks{On leave of absence from Yerevan Physics Institute. }, 
N.~Bianchi$^a$, 
E.~De~Sanctis$^a$, 
W.-D.~Nowak\address{DESY Zeuthen, D-15738 Zeuthen, Platanenallee 6, Germany}
}

\begin{document}

\begin{abstract}
We consider the {\it sin}$\phi$ single 
target-spin asymmetry in deep-inelastic $\pi^{+}$ inclusive 
electroproduction off a longitudinally polarized target.  
We show that at larger transverse momentum of the outgoing hadron the 
evaluated asymmetry 
decreases if one takes into account the first order $\alpha_S$ 
perturbative contribution to the cross section, integrated over the 
azimuthal angle. This leads to good agreement with recent HERMES data. 
\end{abstract}

\maketitle

\section{Introduction}
Single spin asymmetries (SSA) in semi-inclusive deep 
inelastic scattering (SIDIS) of leptons off polarized target are forbidden 
in the simplest version of 
the parton model and in leading order $\alpha_S$ perturbative quantum 
chromodynamics (pQCD);   
they vanish in models based on hadrons consisting of non-interacting 
collinear partons (quarks and gluons). 
Effects due to finite quark transverse momentum play a significant role 
in most explanations of non-zero single-spin asymmetries in polarized 
hadronic reactions~\cite{COL,AK,TM}. Especially polarized fragmentation 
functions with 
transverse momentum dependence are relevant for the description of 
single-target spin asymmetries observed by HERMES~\cite{HERM} and 
SMC~\cite{SMC}. These functions do not vanish even if 
time reversal invariance is applied, because of final-state 
interactions between the outgoing hadron and other produced particles.       

In this note we calculate the dependence of the {\it sin}$\phi$ single spin 
asymmetry of inclusive pion production in $e \vec{p}$-scattering on 
the transverse momentum of the detected hadron, $P_{hT}$. Taking into 
account the first order $\alpha_S$ perturbative contribution to the cross 
section, integrated over 
the azimuthal angle, leads to a good description of the HERMES 
data~\cite{HERM} over the measured $P_{hT}$ range.     
  \section{Formulae}
The kinematics of deep-inelastic pion inclusive 
electroproduction is the following: 
$k_1$ ($k_2$) is the 4-momentum of the 
incoming (outgoing) charged lepton, $Q^2=-q^2$, where $q=k_1-k_2$,  
is the 4-momentum of the virtual photon. 
$P$ ($P_h$) is the momentum of the target (observed hadron), $x=Q^2/2(Pq)$, 
$y=(Pq)/(Pk_1)$, $z=(PP_h)/(Pq)$, $k_{1T}$ is the incoming lepton
transverse momentum with respect to the virtual photon momentum direction, 
and $\phi$ is the azimuthal angle between $P_{hT}$ and $k_{1T}$ around the 
virtual photon direction. Note that the azimuthal angle of the transverse 
(with respect to the virtual photon) component of the target polarization, 
$\phi_S$, is equal to 0 ($\pi$) for the target polarized parallel 
(anti-parallel) to the beam \cite{OABK}. 

The {\it sin}$\phi$ moment in the semi-inclusive deep inelastic cross 
section, which measures the left-right asymmetry of $P_{hT}$ along the 
$k_{1T}$ direction, is given 
by
\begin{equation}
\label{def}
\langle \sin\phi \rangle = \frac{\int d\sigma^{(NP)}\sin\phi}
{\int d\sigma^{(NP)} + \int d\sigma^{(\alpha_S)}},
\end{equation}
where $d\sigma^{(NP)}$ and $d\sigma^{(\alpha_S)}$ are the non-perturba- 
\\tive and the first order $\alpha_S$ perturbative hadronic scattering 
cross sections, respectively. Note, that the leading perturbative effects 
in the SSA appear only at second order $\alpha_S$ pQCD which is beyond of 
the scope of this work.   

The integrations in Eq.(\ref{def}) are over $P_{hT}$, $\phi$,  
$x$, $y$, $z$. The azimuthal asymmetry defined above is related to 
the ones measured by HERMES~\cite{HERM} through the following relation:  
\begin{equation}
 A^{\sin\phi}_{UL}=2 \langle \sin \phi \rangle, 
\label{REL1}
\end{equation}
where the subscripts $U$ and $L$ indicate unpolarized beam and longitudinally 
polarized target, respectively.

The relevant processes for the first order in $\alpha_S$ are the 
subprocesses $e(k_1)+i(p_1) \to e(k_2)+j(p_2)+X$, where $i (j)$ denotes 
initial (final) quarks/gluons~\cite{MEND,CES}, $p_1$ ($p_2$) is the 
incident (scattered) parton momentum and 
\begin{eqnarray}
\label{ALS}
\int d\sigma^{(1)} d\phi &=& \frac{4\alpha_S\alpha^2}{3Q^2} 
{1 \over y} \int_{x}^1 {dx_p \over x_p} 
\int_{z}^1 {dz_p \over z_p} \sum_q e^2_q 
\nonumber \\ 
&& \mbox{} 
\Biggr [ f^q_1(\xi) A D^q_1(\xi^{'})
+f^q_1(\xi) B D^g(\xi^{'})
\nonumber \\ 
&& \mbox{}  
+ f^g_1(\xi) C D^q_1(\xi^{'}) \Biggr ],
\end{eqnarray}
where $f^q_1(\xi)$ is the probability distribution describing a parton 
$q$ with a fraction $\xi$ of the target momentum, $p_1^{\mu}=
\xi P_{1}^{\mu}$, $D^q_1(\xi^{'})$ is the probability distribution for  
a parton $q$ to fragment producing a hadron with a fraction $\xi^{'}$ of 
the parton's momentum, and $e^2_q$ is the charge square of a  
parton $q$.   
In Eq.(\ref{ALS})
\begin{eqnarray}
\label{COF1}
A &=& [1+{(1-y)}^2] \frac{x^2_p+z^2_p}{(1-x_p)(1-z_p)} + 
\nonumber \\ 
&& \mbox{}
2y^2(1+x_pz_p)+4(1-y)(1+3x_pz_p) 
\end{eqnarray}
\begin{eqnarray}
\label{COF2}
B &=& [1+{(1-y)}^2]\frac{x^2_p+({1-z_p})^2}{z_p(1-x_p)}
\nonumber \\ 
&& \mbox{}
+2y^2(1+x_p-x_pz_p)
\nonumber \\ 
&& \mbox{}
+4(1-y)(1+3x_p(1-z_p)) 
\end{eqnarray}
\begin{eqnarray}
\label{COF3}
C &=& {3\over8} \Biggr[[1+{(1-y)}^2]
\nonumber \\
&& \mbox{}
[x^2_p+({1-x_p})^2]\frac{z^2_p+({1-z_p})^2}
{z_p(1-z_p)}
\nonumber \\ 
&& \mbox{}
+16x_p(1-y)(1-x_p) \Biggr ],
\end{eqnarray}
and $x_p$ and $z_p$ are the parton 
variables which are related to the hadron variables as
$ x_p=x/\xi=Q^2/2p_1q, z_p=z/\xi^{'}=p_1p_2/p_1q $. 
These expressions are identical to previous perturbative results in 
Ref.\cite{MEND,CES,OABD}. 

The $P_{hT}$-integration of the non-perturbative cross section can be 
performed analytically assuming that the transverse momentum 
dependence in the distribution and fragmentation functions can be 
written in factorized exponential form:  
\begin{equation}
\label{gauss1}
 f(\vec{p}_T) = {1 \over a^2\pi} e^{-p_T^2/a^2} 
\end{equation}
\begin{equation}
\label{gauss2}
 d(\vec{k}_T)={z^2 \over b^2\pi} e^{-{z^2 k_T^2}/b^2},    
\end{equation}
where $p_T$, $k_T$ are the intrinsic transverse momenta of the 
initial and final quark, respectively and $a=2 \langle 
p_T \rangle /\sqrt{\pi}, 
b=2 \langle z k_T \rangle /\sqrt{\pi}$. 
Then (for more details see Refs. \cite{TM,OABK}) 
\begin{eqnarray}
\int d{\sigma}d^2P_T &=& S(P_C)
   (1+(1-y)^2)f_1(x) D_1(z), \label{FD1} \\
   \int d{\sigma} \sin \phi d\phi &=&  
S(P_{hT}) \biggl \{ S_L
2(2-y) \sqrt{1-y}{P_{hT} \over {z Q}} \nonumber \\ 
&& \biggl [ R_1 x h_L (x) H_1^{\perp}(z) - \nonumber \\
&& R_2 g_1(x) H_1^{\perp}(z) - 
R_3 h_{1L}^{\perp}(x) {\tilde{H}(z) \over z}\biggr ] + \nonumber \\
&& S_{Tx}(1-y){P_{hT}\over z}R_4h_1(x)H_1^{\perp}(z) 
\biggr \}\, \label{GHT}. 
\end{eqnarray}
Here 
$$
S(P)={4\pi^2 \alpha^2 \over Q^2 y} \exp(-\frac{P^2}{b^2+a^2z^2})\, 
, 
$$
and 
$$
R_1 = {M_pb^2 \over M_h {(b^2+a^2z^2)}^2}, \quad 
R_2 = {mb^2 \over M_h{(b^2+a^2z^2)}^2}, 
$$
$$ 
R_3 = {M_h a^2 z^2 \over M_p {(b^2+a^2z^2)}^2}, \quad 
R_4 = {b^2 \over M_h {(b^2+a^2z^2)}^2}.
$$
where $M_p$, $M_h$, and $m$ are the proton, final hadron and current quark 
masses, respectively. In Eq.(\ref{GHT}) the components of the 
longitudinal and transverse target polarization 
in the virtual photon frame are denoted by $S_L$ and $S_{Tx}$, 
respectively~\cite{OABK}. Note that to get rid off the divergences 
at $P_{hT}$ close to zero ($x_p \to 1$, $z_p \to 1$) and to control the 
integration of the order-$\alpha_S$ cross section~\cite{CES},  
we introduce $P_C$ as a lower cutoff for $P_{hT}$. This has no impact on 
our results because of we are not considering the fully inclusive 
cross section. We perform all numerical investigations for a series of 
$P_C$ cutoffs, which approximates the $P_{hT}$ dependence of the asymmetry. 
Twist-2 distribution and fragmentation functions have a subscript `1':
$f_1(x)$ and $D_1(z)$ as already mentioned above are the usual unpolarized 
distribution and fragmentation functions, $g_1(x)$ is the longitudinally 
polarized distribution function, while 
$h^{\perp }_{1L}(x)$ and $h_1(x)$ describe the quark transverse spin 
distribution in longitudinally and transversely polarized nucleons,  
respectively. The interaction-dependent part of the twist-3 distribution 
function in the longitudinally polarized nucleon, $h_L(x)$~\cite{JJ91}, is 
denoted by $\tilde{h}_L(x)$~\cite{JJ92,MT}. The spin-dependent 
twist-2 fragmentation 
function $H_1^{\perp }(z)$, describing 
transversely polarized quark fragmentation~\cite{COL},  
correlates the transverse spin of a quark with a preferred transverse 
direction for the production of the pion.  
The fragmentation 
function $\tilde{H}(z)$ is the interaction-dependent part of the twist-3 
fragmentation function~\cite{TM}.
     
\section{Numerical Results}
For numerical calculations the non-relativistic approximation $h_1(x) = 
g_1(x)$ is used as a lower limit~\footnote{In the non-relativistic quark
model $h_1(x,\mu^2_0) = g_1(x,\mu^2_0)$. Several models suggest that  
$h_1(x)$ has the same 
order of magnitude as $g_1$~\cite{JJ92,PP,BCD}. The evolution properties 
of $h_1$ and 
$g_1$, however, are very different~\cite{SV}. At the $Q^2$ values of the 
HERMES measurement 
the assumption $h_1=g_1$ is fulfilled at large $x$ values (valence-like 
region), 
while large differences occur at lower $x$~\cite{KNO}. }, 
and $h_1(x)=(f_1(x)+g_1(x))/2$ as an upper limit~\cite{SOF}.  
For the sake of simplicity, $Q^2$-independent parameterizations were chosen  
for the distribution functions $f_1(x)$ and $g_1(x)$~\cite{BBS}. We use the 
approximation where the twist-2 {\it transverse} 
quark spin distribution in the {\it longitudinally} polarized nucleon 
is zero (i.e. $h_L(x)=\tilde{h}_L(x)=h_1(x)$) which, as shown in 
Ref.\cite{DNO,OBDN}, leads to a consistent description 
of the $A_{UL}^{sin\phi}$ and $A_{UL}^{sin2\phi}$ asymmetries observed 
by the HERMES collaboration~\cite{HERM}. 
\begin{figure}[ht]
\begin{center}
\includegraphics[width=7.5cm]{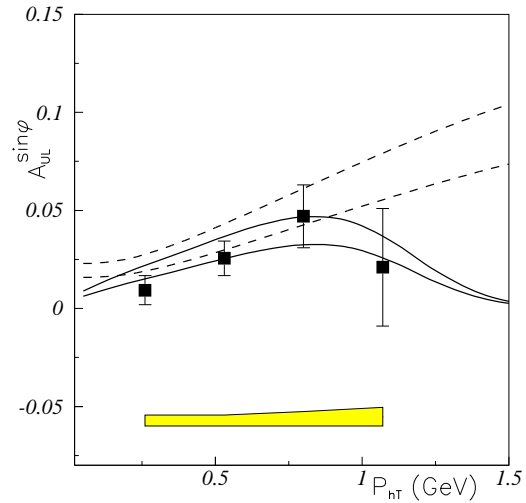}
\caption{$P_{hT}$ dependence of the $A^{\sin\phi}_{UL}$ asymmetry for 
$\pi^{{+}}$ electroproduction off longitudinally polarized protons  
evaluated using $M_C=2 m_{\pi}$ and $\eta=0.6$ in Eq.(\ref{H1T}). 
Full lines correspond to results where the perturbative contribution 
is taken into account in the denominator of Eq.(\ref{def}), while dashed 
ones are without this contribution. For each case two curves are presented 
corresponding to $h_1=g_1$ (lower curve) and $h_1=(f_1+g_1)/2$ (upper curve).
HERMES data are from Ref.~\cite{HERM}. }
\label{pc}
\end{center}
\end{figure}
To estimate the T-odd fragmentation function $H_1^{\perp }(z)$,
the Collins type ansatz~\cite{COL} for the analyzing power of 
transversely polarized quark fragmentation was adopted:
\begin{equation}
A_C(z,k_T) \equiv \frac{\vert k_T \vert}{M_h}\frac{H_1^{\perp}(z,k_T^2)}
{D_1(z,k_T^2)} = \eta \frac{M_C\,\vert k_T \vert}{M_C^2+k_T^2}.
\label{H1T}
\end{equation}
Here $\eta$ is taken as a constant, although in principle it could be $z$ 
dependent and $M_C$ is a typical hadronic mass whose value ranges from
$2m_{\pi}$ to $M_p$. 
 
For the distribution of the final parton's intrinsic transverse momentum, 
$k_T$, in the unpolarized fragmentation function $D_1(z,k^2_T)$ a Gaussian 
parameterization was used~\cite{KM} with $\langle z^2 k_T^2 \rangle = b^2$ 
(in the numerical calculations $b = 0.36$ GeV was taken~\cite{PYTHIA}). 
For $D_1^{\pi^{+}} (z)$ the parameterization from Ref.~\cite{REYA} was 
adopted.

In Fig.~\ref{pc}, the asymmetry $A^{\sin\phi}_{UL}$ 
of Eq.(\ref{REL1}) for $\pi^{+}$ production on a proton target is presented 
as a function of $P_{hT}$ and compared to the HERMES data~\cite{HERM}. 
The results obtained with and without 
taking into account the leading order $\alpha_S$ pQCD term in the 
denominator of 
Eq.(\ref{def}), are denoted by pairs of full and dashed lines, respectively. 
Each pair of curves corresponds to the two limits 
chosen for $h_1(x)$. From Fig.~\ref{pc} it can be seen that  
taking into account the first order $\alpha_S$ perturbative 
contribution to the cross section (integrated over the azimuthal angle), 
the asymmetry $A^{\sin\phi}_{UL}$ goes down at higher $P_{hT}$. 
This leads to a good agreement with HERMES data~\cite{HERM}. 

\section{Conclusion}

We have discussed  the $P_{hT}$ behavior of the {\it sin}$\phi$  
single-spin asymmetry for $\pi^{+}$ production in semi-inclusive deep 
inelastic scattering of leptons off longitudinally polarized protons 
including the pQCD contribution to the $\phi$-independent cross section. 
We have shown that the HERMES data at larger $P_{hT}$ can be 
described well if one takes into account the first order $\alpha_S$
contribution. 

\section{Acknowledgments}

We thank D.~Boer, A.~Brandenburg and R.L.~Jaffe for interesting discussions. 
The work of K.A.O. was in part supported 
by INTAS contributions (contract numbers 93-1827 and 96-287) 
from the European Union.

\end{document}